\documentclass[a4paper]{PoS}

\DeclareGraphicsExtensions{.jpg,.mps,.pdf,.png,.eps} 
\usepackage{epstopdf}
\usepackage{wrapfig}

\title{Muon tomography applied to active volcanoes}

\ShortTitle{Diaphane project}

\author{\speaker{Jacques Marteau}\\
Institut de Physique Nucl\'eaire de Lyon, Univ Claude Bernard, UMR 5822 CNRS, Lyon, France\\
\email{marteau@ipnl.in2p3.fr}}
	
\author{Bruno Carlus\\
Institut de Physique Nucl\'eaire de Lyon, Univ Claude Bernard, UMR 5822 CNRS, Lyon, France}	

\author{Dominique Gibert\\
OSUR - G\'eosciences Rennes, Univ Rennes 1, UMR 6118 CNRS, Rennes, France\\
Volcano Observatories, Institut de Physique du Globe de Paris, Paris, France}

\author{Jean-Christophe Ianigro\\
Institut de Physique Nucl\'eaire de Lyon, Univ Claude Bernard, UMR 5822 CNRS, Lyon, France}	

\author{Kevin Jourde\\
Institut de Physique du Globe de Paris, Sorbonne Paris Cit\'e, UMR 7154 CNRS, Paris, France}

\author{Bruno Kergosien\\
OSUR - G\'eosciences Rennes, Univ Rennes 1, UMR 6118 CNRS, Rennes, France}

\author{Pascal Rolland\\
OSUR - G\'eosciences Rennes, Univ Rennes 1, UMR 6118 CNRS, Rennes, France}

\abstract{Muon tomography is a generic imaging method using the differential absorption of cosmic muons by matter. The measured contrast in the muons flux reflects the matter density contrast as it does in conventional medical imaging. The applications to volcanology present may advantadges induced by the features of the target itself: limited access to dangerous zones, impossible use of standard boreholes information, harsh environmental conditions etc. The Diaphane project is one of the largest and leading collaboration in the field and the present article summarizes recent results collected on the Lesser Antilles, with a special emphasis on the Soufrière of Guadeloupe.}

\FullConference{International Conference on New Photo-detectors,PhotoDet2015\\
		6-9 July 2015 \\
		Moscow, Troitsk,Russia}

\begin{document}

\section{Introduction}

\noindent One of the first applications of muon tomography is due to Alvarez in his attempt to scan the Chephren pyramid and look for hidden chambers \cite{Alvarez:1970}. The unconclusive results were mainly due to technical limitations of the detectors used at that time. The same principles were then used for volcano radiography first in Japan \cite{nagamine1995geo, nagamine1995method, tanaka2001, tanaka2005}, then in Europe \cite{lesparre2010, lesparre2012a, lesparre2012b, carloganu2012}. The societal implications of such research activities is obvious since most of the volcanoes are close to populated areas. Therefore the monitoring of their activity, the understanding of their behaviour and the evaluation of the associated risks for surrounding inhabitants are of prime interest. This requires accurate imaging of the volcano's dome and quantitative estimates of the mass distributions and of the associated fluid transports (magma, gas or water).  

\begin{figure}[!ht] 
   \centering
   \includegraphics[width=4.5cm,height=4.5cm]{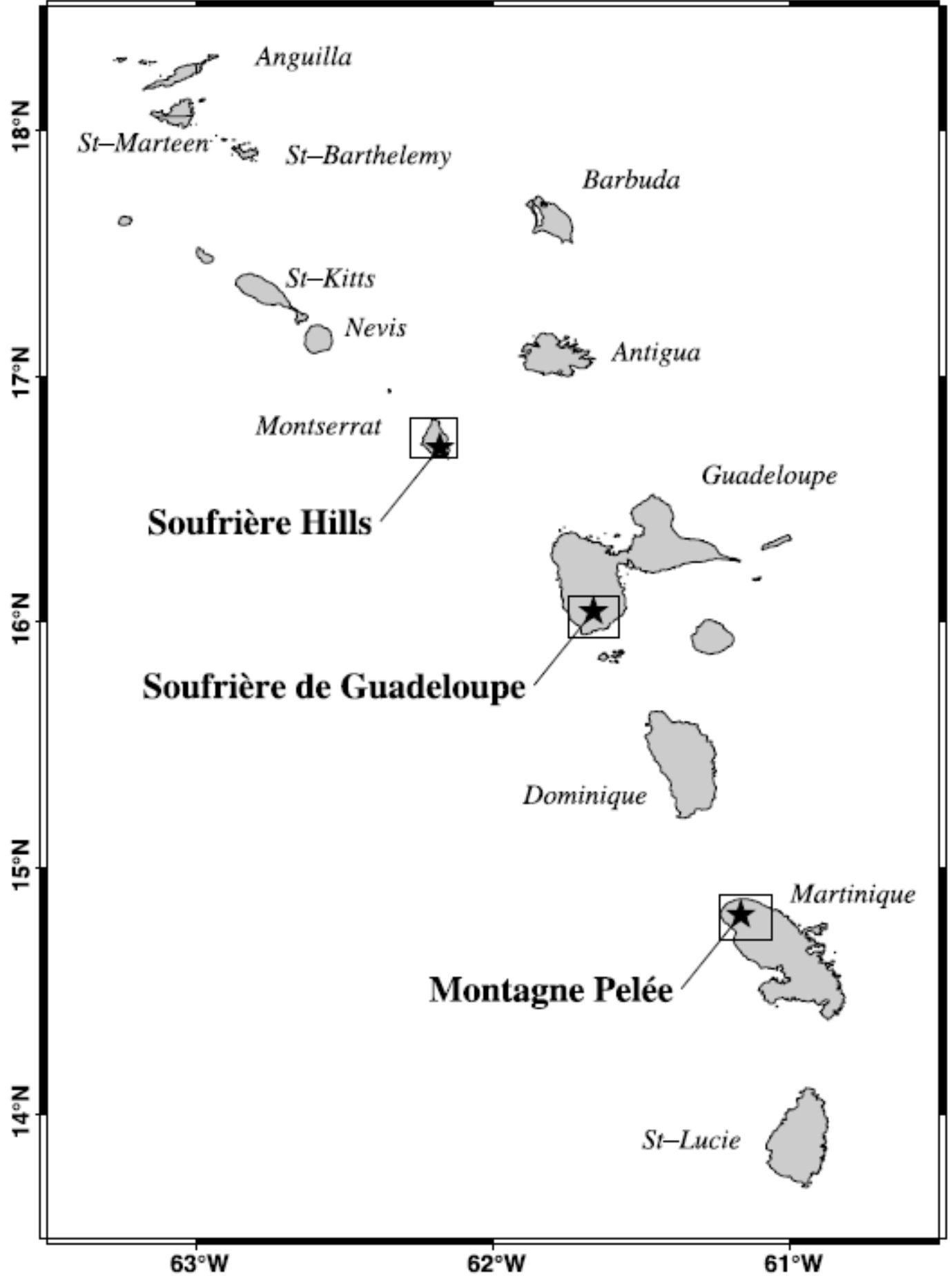} \hfill
   \includegraphics[width=4.5cm,height=4.5cm]{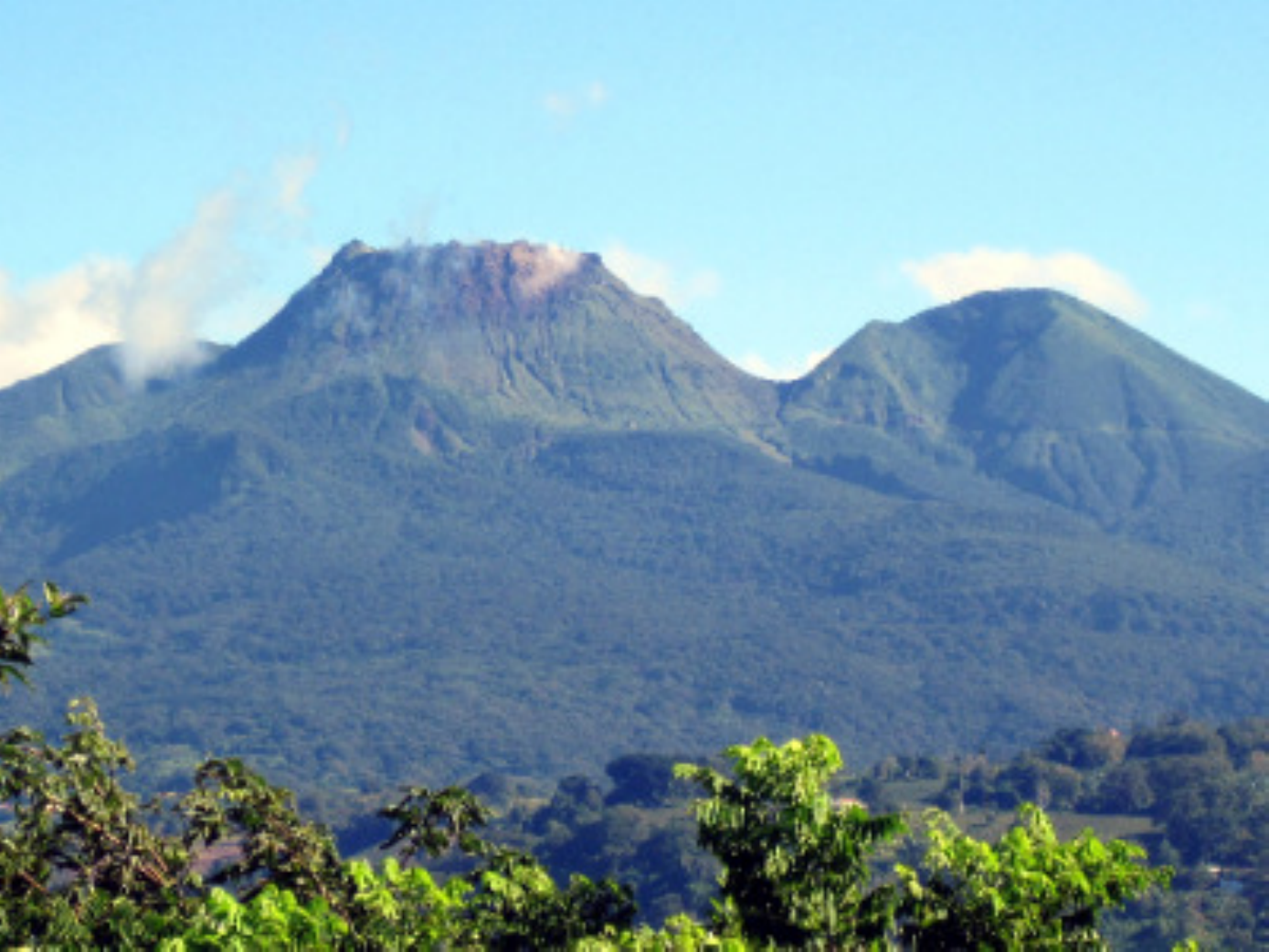} \hfill
   \includegraphics[width=5cm,height=4.5cm]{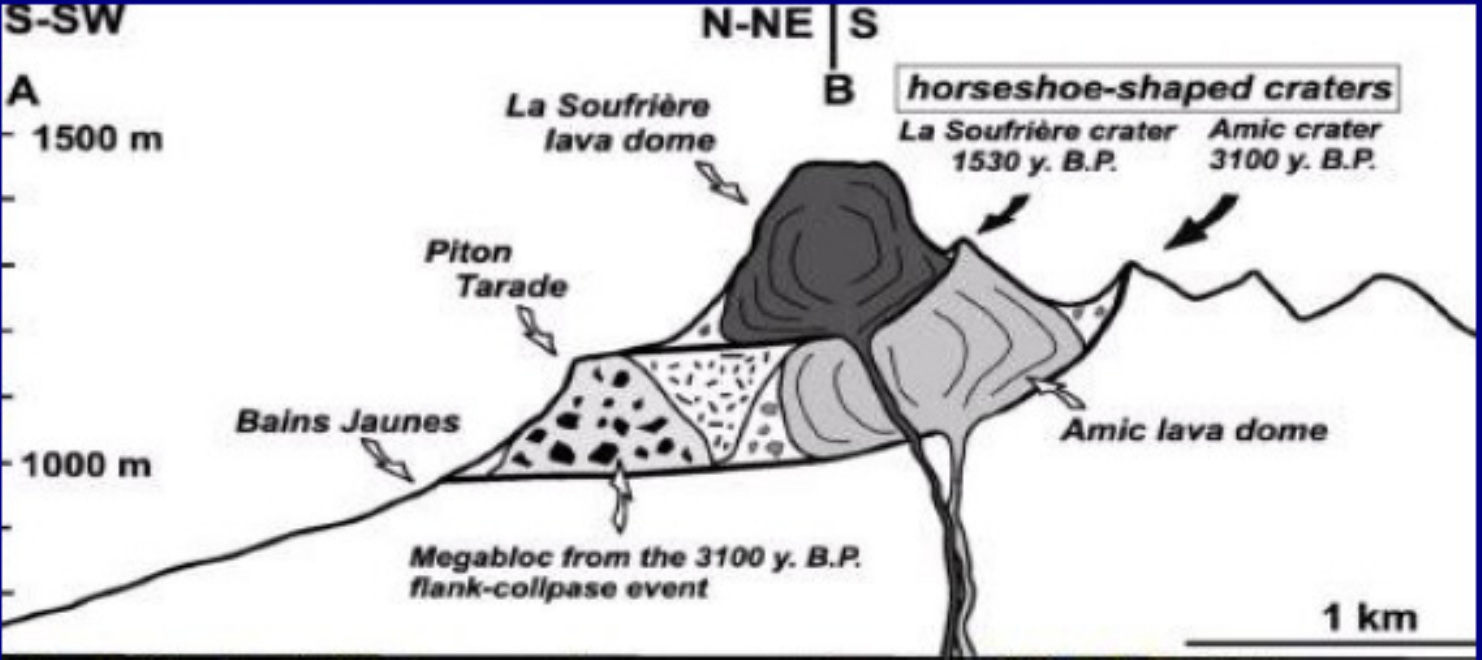} 
   \caption{Location of Lesser Antilles islands and active volcanoes (left). La Soufri\`ere of Guadeloupe: view from the observatory (middle) and geological model (right).}
   \label{Fig:Antilles}
\end{figure}

\noindent DIAPHANE (IPG Paris, IPN Lyon and G\'eosciences Rennes) is the first european project of tomography applied to volcanology and underground structures characterization. The volcanologic part of the project focuses on the Lesser Antilles, a subduction volcanic arc with a dozen of active volcanoes, such as the Montagne Pel\'ee in Martinique, the Soufri\`ere of Guadeloupe, and Soufriere Hills in Montserrat, which all presented eruptive activity during the 20$^\mathrm{th}$ century. Fig.\ref{Fig:Antilles}, left, shows the location of those volcanoes. In particular the Soufri\`ere of Guadeloupe (Fig.\ref{Fig:Antilles}, middle) is an andesitic volcano which lava dome is about five hundred years old \cite{Boudon:2008, Komorowski:2008}. Its dome sits on a 15$^o$ N-S inclined plane (Fig.\ref{Fig:Antilles}, right) leading to an unstable structure and is very heterogeneous, with massive lava volumes embedded in more or less hydrothermalized materials \cite{Nicollin:2006, lefriant2006}. Given the constant erosion of the volcano due to the tropical intensive rain activity, the evolution of such a lacunary structure may be rapid, with cavities filled with pressurized and likely acid fluids. These features, which are common to many tropical volcanoes, lead to the choice of the Soufrière of Guadeloupe as a priority target \cite{gibert2010}.

\section{Tomography basics and results}

\begin{wrapfigure}{r}{0.7\textwidth} 
   \centering
   \includegraphics[width=0.7\textwidth]{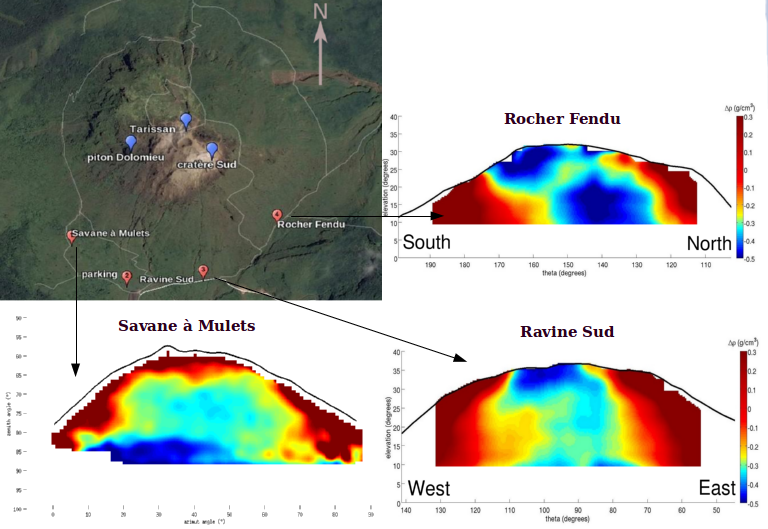} 
   \caption{Density profiles obtained from 2010 to 2015 at three different locations around the Soufrière of Guadeloupe (points 1, 3, 4 on the map).}
   \label{Fig:TomoSouf}
\end{wrapfigure}

\noindent Muon radiography proceeds like standard medical imaging by measuring the attenuation of a beam (cosmic muons) which crosses matter (e.g. the dome of a volcano) with a sensitive device (so-called muon ``telescope'') \cite{lesparre2010}. The measurement gives access to the opacity of the structure $\varrho$, which is the integral of the density along the muon trajectory in the matter, by comparing the muons flux $\Phi$ after crossing the target, to the incident open sky flux, $\Phi_o$. Various models give analytical expressions of the muon flux from the two-body decays of pions and kaons produced by interactions of the incident primary protons \cite{Bugaev:1998, gaisser2008}. The presence of matter on the trajectory acts as a filter  since only the most energetic muons will escape the structure. 
The muons energy loss on their way through rock, $-dE/dx$, accounts for the standard physical processes\footnote{For a review of those processes, see http://pdg.lbl.gov}. The flux of muons emerging from the target is influenced by environmental parameters such as altitude, geomagnetic cut-off, solar modulation, atmospheric variations\footnote{An example of barometric corrections is given in Section 5.} to be accounted for in the simulation models. 
\noindent The last step of the analysis is the inversion of the problem to go from the measured attenuated muon flux to the density $\rho$ maps e.g. in the (azimut angle, zenith angle). Typical examples of such maps are given in Fig.\ref{Fig:TomoSouf}. The quality of the obtained images has been improved by a time-of-flight -- tof -- analysis which rejects particles propagating backwards and mimicking particles emerging from the volcano with nearly horizontal trajectories, that is from the regions with the largest opacity \cite{jourde2013}. This tof analysis is possible thanks to a TDC vernier-technique coded in a FPGA, with no hardware modification to the original design \cite{marteau2014}. The accuracy on the timestamping has been improved to $250$ps steps, which is sufficient for the sampling of the typical ns-scale muons tof and for an average background substraction of the wrong-direction trajectories.\\

\noindent The density maps shown in Fig.\ref{Fig:TomoSouf} reflect the complex inner structure of the volcanic dome. The three views obtained exhibit not only a very good compatibility with each other but also with other measurements carried out with different methods on the same place (gravimetry and electrical tomography \cite{lesparre2012c}. The density maps indicate presence of large low density volumes within the cone, also seen in electrical tomographic data (highly conductive zones being inferred either to hydrothermally washed zones or to acid zones), and reveals the existence of large hydrothermal channels to be accurately monitored. These 2D views are being combined to get real-time 3D analysis from the Soufrière dome.      

\section{Photo-active detectors for tomography}

\noindent The design of the DIAPHANE telescopes relies on well-established and robust detector technology~: plastic scintillator with WLS fibres, multi-anode PMT's (Hamamatsu 64 channels H8804-type) or MPPC's (S10362-11-050C) and triggerless, smart, Ethernet-capable R/O electronics, based on the state-of-the-art opto-electronics technology. The detectors design should cope with the most stringents constraints in terms of autonomy, power consumption (less than 50W in total), mass, remote accessibility etc imposed by transportation restrictions and harsh environmental conditions \cite{lesparre2010}. Each detection matrices is made of two scintillator bars layers ($X$ \& $Y$). Two types of scintillators are being used~: $5 \times 1 \; \mathrm{cm}^2$ Fermilab bars co-extruded with a TiO$_2$ reflective coating and a central fibre groove or $2,5 \times 0,7 \; \mathrm{cm}^2$ JINR-type bars painted and with a surface groove. At least 3 matrices are used in coincidence in a complete telescope to reject random coincidences. The global DAQ system is a reduced version of the OPERA one \cite{Marteau:2009} and built as a network of ``smart sensors'', synchronized by a common GPS-disciplined clock unit.\\

\begin{wrapfigure}{r}{0.5\textwidth}
 \centering 
 \includegraphics[width=0.5\textwidth]{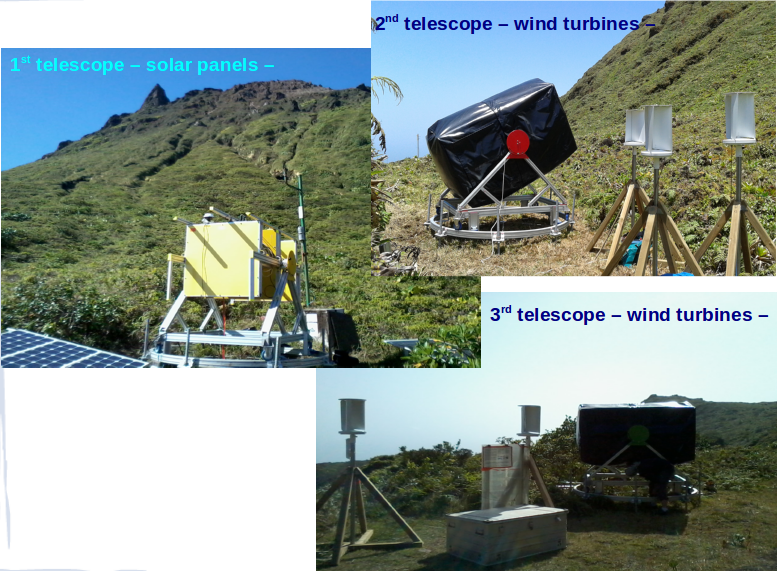} 
 \caption{Pictures of the three Soufri\`ere telescopes (yellow: ``Parking'', red: ``Roche fendue'', green:``Matylis'').}
 \label{Fig:Telescope}
\end{wrapfigure}

\noindent Seven Diaphane telescopes have been recording data on active volcanoes (Mount Etna in Sicily \cite{carbone2014}, the Mayon in the Philippines and the Soufrière in Guadeloupe) or in undergound laboratories (the Mont-Terri in Switzerland \cite{BossartThury:2008} and the Tournemire laboratory in France). The underground experiments allowed to validate the detectors design and to measure their performance. Three telescopes are now running and will be moved around the Soufrière volcano to perform real-time 3D scans of the dome. A smaller telescope has been built and put in a fault under the South crater to monitor directly the zone beyond the most active part of the dome. Muons telescopes will run on-site until 2018.

\section{Coupling with other methods}

\noindent We investigated also, in a resolving kernels approach, how the resolution of small-scale geological density models is improved with the fusion of information provided by gravity measurements and muon radiographies \cite{jourde2015}. The two methods differ significantly since one involves straight-ray transmission while the other is a 3D-integrative method. The sensitive regions are also different, muon tomography seeing only "above the horizon" while gravimetry brings information on the whole density distribution.Preliminary works on the subject remain very scarce (see \cite{jourde2015} and references therein). 

\begin{figure}[!ht] 
 \centering
 \includegraphics[width=0.85\textwidth]{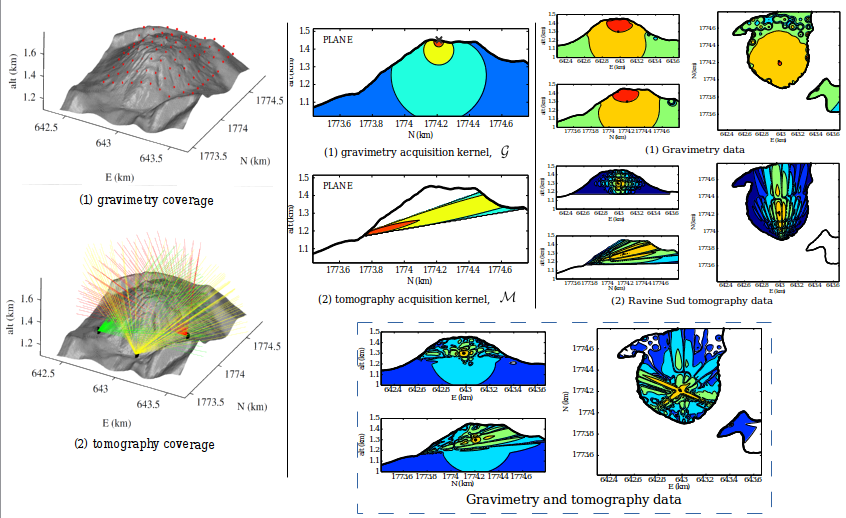} 
 \caption{Muon-gravimetry joined analysis. Left: data coverage for gravimetry (top) and muon tomography (bottom). Middle: acquisition kernels for both methods (same convention). Right: separate resolving kernels (top and middle for gravimetry and muon tomography, bottom for the joined analysis).}
 \label{Fig:Muongravi}
\end{figure}

\noindent The resolving kernels derived in the joined muon/gravimetry showed interestingly that the resolution in deep regions not sampled by muon tomography is significantly improved by joining the two techniques as illustrated in Fig.\ref{Fig:Muongravi} on the example of the Soufrière. The measurement points and data coverage for both methods is displayed on the left. On the left we show the acquisition and the resolving kernels for both methods separated. The dashed plots show the joined gravimetry plus muon tomography resolving kernel. The improvement of the resolution on the deepest regions, where no direct information is brought by muon tomography, is clear. This method is being currently systematically used on the field. 

\section{Methodologic developments and conclusions}

\begin{wrapfigure}{r}{0.6\textwidth}
 \centering
 \includegraphics[width=0.6\textwidth]{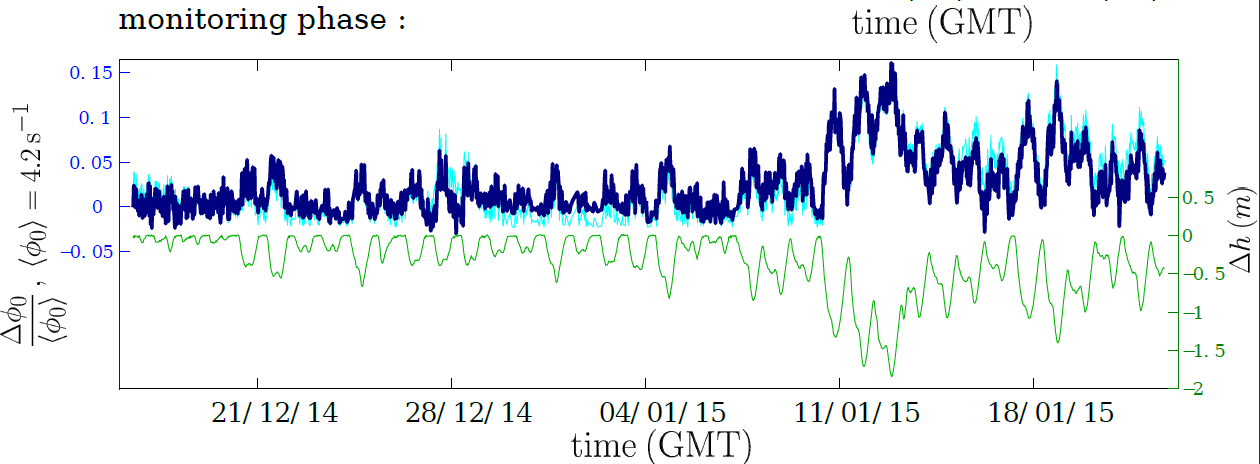}
 \caption{Normalized and centred muon flux (light blue: raw data, dark blue: barometric corrections applied) plus tank water level (green) changes as a function of time.}
 \label{muon_flux_and_water_level_figure}
\end{wrapfigure}

\noindent The so-called SHADOW experiment is a methodological development (November 2014 - February 2015) where we put a muon telescope below the tank of a domestic water-tower. The muon flux changes are correlated with the variations of the water level in the tank since the water level directly governs the opacity. Muons flux is continuously measured by the telescope and correlated with the measurements obtained with a standard gauge. The goal of the experiment is not only to assess the dynamical sensitivity of the method but also the sizes of higher level corrections such as atmospheric pressure (proportionnal in a first approximation to the atmospheric opacity) and temperature. Fig.~\ref{muon_flux_and_water_level_figure} displays, as a function of time, the monitored water level (green curve), and the measured muons flux without (light blue) and with barometric correction (dark blue). 

\noindent During the beginning of the data taking, the water level was almost constant with a reference maximal height of $4.96$m. This first period is used as a reference to measure the barometric corrections. After this first data taking period, regular water level variations occurred, up to 50\% relative change in the maximal height. The muons data show the expected  anti-correlation between the water level and the measured flux.   
The anti-correlation parameter is extracted from a fit of the relative flux variation $\Delta\Phi_0 / \langle \Phi_0 \rangle$ versus the relative height $\Delta h$. The barometric correction clearly improves the quality of the fit and appears not to be negligible in this type of measurements with low opacity targets, sensitive to low energy muons.\\ 

\noindent This small experiment reinforced our belief that the method is not only capable of performing static structural images, but also monitoring the dynamics of the hydrothermal system of a tropical volcano. Indeed during the 2012 measurement campaign on the Soufrière, it has been noticed that there were significant changes in the muons flux. Those changes could not be correlated with any instrumental effects and they coincided with the appearance of new vents at the summits of the volcano. A possible interpretation, still under investigation, is the vaporization of the water inside the hydrothermal system, which leads to a smaller opacity within the dome and therefore a larger muon flux. The monitoring of density time-changes is being analyzed in more details. \\

\noindent R\&D around the muon tomography method shows that the geosciences applications are rich, once the adaptation of the detectors to the environmental conditions is under control. Robust and ``unmanned'' autonomous detectors have been design and built by the DIAPHANE collaboration and are now operating on various active volcanoes, bringing a significant amount of new information to better constrain the existing models.

\end{document}